\documentclass[aps,showpacs,preprintnumbers,amsmath,amssymb]{revtex4}

\usepackage{float}
\usepackage{graphics,epsfig}
\usepackage{graphicx}
\usepackage{dcolumn}
\usepackage{bm}
\usepackage{color}

\begin{document}
\baselineskip=0.8 cm
\title{{\bf  Quasinormal modes of black holes absorbing dark energy }}

\author{Xi He$^{1}$, Bin Wang$^{1}%
$\footnote{wangb@fudan.edu.cn }, Shao-Feng Wu
$^{2}$\footnote{sfwu@shu.edu.cn}, Chi-Yong
Lin$^{3}$\footnote{lcyong@mail.ndhu.edu.tw}} \affiliation{$^{1}$
Department of Physics, Fudan University, 200433 Shanghai}
\affiliation{$^{2}$ Department of Physics, Shanghai University,
200436, Shanghai } \affiliation{$^{3}$ Department of Physics,
National Dong Hwa University, Shoufeng, 974 Hualien}

\vspace*{0.2cm}
\begin{abstract}
\baselineskip=0.6 cm
\begin{center}
{\bf Abstract}
\end{center}
We study perturbations of black holes absorbing dark energy. Due to
the accretion of dark energy, the black hole mass changes. We
observe distinct perturbation behaviors for absorption of different
forms of dark energy onto the black holes. This provides the
possibility of extracting information whether dark energy lies above
or below the cosmological constant boundary $w=-1$. In particular,
we find in the late time tail analysis that, differently from the
other dark energy models, the accretion of phantom energy exhibits a
growing mode in the perturbation tail. The instability behavior
found in this work is consistent with the Big Rip scenario, in which
all of the bound objects are torn apart with the presence of the
phantom dark energy.

\end{abstract}

 \maketitle

\newpage

There has been growing observational evidence showing that our
universe is accelerated expanding driven by a yet unknown dark
energy (DE) \cite{obsI,obsII,obsIII}. The leading interpretation of
such dark energy is a cosmological constant with equation of state
$w=-1$. More sophisticated models have been proposed to replace the
cosmological constant by either relating the dark energy to a scalar
field called quintessence with $w>-1$, or to an exotic field called
phantom with $w < -1$. But it is doubtful that there is a clear
winner in sight to explain the nature of dark energy at the moment.
Recently, extensive analysis found that the current data favors dark
energy models with equation of state in the vicinity of $w =-1$
\cite{cald}, straddling the cosmological constant boundary. This
observational value was pinned down through large scale surveys from
CMB, large scale structure and SNIa observations.

At present the equation of state is the only antenna to learn the
microscopic nature of dark energy. Its observational value is
particularly important to determine whether the dark energy is of
the phantom type, quintessence type or cosmological constant.
Besides the available observational methods, it is of great interest
to devise other complementary tools to measure the values of
$\omega$. In this work we use quasinormal modes (QNM) of a black
hole to investigate the nature of DE equation of state. As is well
known, if the dark energy is modeled as a background cosmic fluid,
the flux of dark energy accreted by the black hole will change its
mass. Now, it is reasonable to take into account these imprints in
perturbations around a black hole and to extract information about
$w$. An attempt in this direction was studied in \cite{laura} by
exploiting the gravitational wave radiated from a binary of
supermassive black holes. It is expected that the binaries observed
with LIGO or LISA can distinguish, in the nearly future, more
accurately values of $w$ of dark energy.

To extract information from the gravitational wave observation and
pin down the value of $w$ of dark energy, we need accurate waveforms
on the perturbations around black holes. There has been great
progress in studying perturbations around black holes recently. In
asymptotically flat spacetimes it is possible to get a schematic
picture regarding the dynamics of waves outside black holes. After
the initial pulse, the perturbation will experience a quasinormal
ringing, which describes the damped oscillations under perturbations
in the surrounding geometry of a black hole with frequencies and
damping times of oscillations entirely fixed by the black hole
parameters. The quasinormal modes is believed as a unique
fingerprint to directly identify the black hole existence. Detection
of these QNM is expected to be realized through gravitational wave
observation in the near future \cite{rev}. At late times,
quasinormal oscillations are swamped by the relaxation process. This
relaxation is the requirement of the black hole no hair theorem
\cite{8}. In Anti-de Sitter and de Sitter spacetimes, perturbations
around black holes have been shown as a theoretical testing ground
to get deeper understandings of the AdS/CFT \cite{3a,3b,4a,4b,4c},
dS/CFT \cite{5} correspondences. More recently, it has been also
argued that perturbations in black hole backgrounds is useful to
extract information on black hole phase transitions
\cite{Shen:2007xk,7}.

In the universe filled with dark energy modeled as scalar field,
the action has the form
\begin{eqnarray}
S=\int d^4x \sqrt{-g}\bigg[{R\over 16\pi G}- {1\over 2}\partial _\mu
\psi \partial^\mu \psi -{1\over 2} \mu^2 \psi^2\bigg].
\end{eqnarray}
where $\mu $ is the mass of the scalar field, and we have taken the
metric sign ($-,+,+,+$). If the dark energy is described by the
phantom field, the kinetic term in the action has the positive sign
\cite{car}. Varying the action with respect to $\psi $, we can
obtain the wave equation in the curved spacetime

\begin{eqnarray}\label{KG}
{1\over \sqrt{-g}} \partial _\mu \big( \sqrt{-g} g^{\mu \nu}
\partial_\nu\pm \mu^2 \psi\big)=0,
\end{eqnarray}
where the $'+'$ sign describes the the phantom field while the
$'-'$ sign describes the quintessence field. The wave equation of
the quintessence field is the same as the massive scalar field
discussed in \cite{mass}.

Considering next the effect of the accretion of dark energy onto the
black hole, the black hole mass changes at a rate \cite{bh}
\begin{eqnarray}\label{Mt}
\frac{dM}{dt} =4\pi A M^2 \rho_\infty (1+w),
\end{eqnarray}
where $A\simeq 5.6 $ is the numerical factor, which define the
energy flux of DE onto the black hole, $\rho_\infty$ is the energy
density of dark energy far away from the hole. The black hole is
no longer static due to the absorption of dark energy, unless the
dark energy being cosmological constant. Accreting the
quintessence field with $w>-1$, the black hole mass increases.
However the black hole mass decreases during the accretion of
phantom energy with $w<-1$. It was argued that the black holes are
not torn apart, but disappear by the Big Rip due to the accretion
of phantom energy \cite{bh}. Other discussions on the change rate
of the black hole mass due to the absorption of dark energy can be
found in \cite{new}. What would be the fate of the perturbation
around black holes absorbing dark energy? To answer this question
we shall employ the formalism first developed in
Ref.\cite{Shao:2004ws}. Some other studies on the time-dependent
background QNMs can be found in \cite{Xue:2003vs}.

To describe the time-dependent black hole background, we start with
the Vaidya metric \cite{Vaidya1953}
\begin{eqnarray}\label{Vaidya}
ds^2=-f dv^2+2c dv dr +r^2 \big(d\theta^2 + \sin^2\theta d\phi^2\big),
\end{eqnarray}
where $f=1-{2 M (v)\over r}$ and $M(v)$ is an arbitrary function
of time. The coordinate $v$ is usually called ``advanced time" and
$c=1$ describes black hole with ingoing radial flow. The horizon
of the Vaidya black hole is inferred from the null hypersurface
condition $g^{\mu \nu} {\partial \tilde{f} \over\partial x^\mu
}{\partial \tilde{f} \over\partial x^\nu }=0$ and
$\tilde{f}(r_+,v)=0$. In this case, $r_+(v)$ satisfies
\begin{eqnarray}\label{horizon}
r_+ -2M(v)-2c r_+ \dot{r}_+=0,
\end{eqnarray}
where $\dot{r}_+={dr_+\over dv}$.

Similarly to the static approach, we introduce the tortoise
coordinate $r_*$ as
\begin{eqnarray}
r_*=r+{1\over 2 \kappa} \ln (r-r_+),
\end{eqnarray}
where $\kappa$ is the surface gravity. Since
$r_*=r_*\big(r,r_+(v)\big), dv=dt+dr_*$, one has
\begin{eqnarray}\label{Mv}
{dM\over dv}&=& \bigg(1-{\partial r_* \over \partial v}\bigg)
{dM\over dt} \\ \nonumber &=& \bigg[1+ {\dot{r}_+ \over 2\kappa
(r-r_+)}\bigg]{dM\over dt}.
\end{eqnarray}

Using the Vaidya metric and the tortoise coordinate, the wave
equation (\ref{KG}) in the time-dependent black hole background can be written
as
\begin{eqnarray}
\bigg[{r^2\over c}{\partial^2  \over \partial v \partial r}+{1\over
c}{\partial \over \partial r}\bigg(r^2 {\partial \over
\partial v}\bigg) + {1\over c^2}{\partial \over \partial r}\bigg(r^2 f
{\partial \over
\partial r}\bigg) +{1\over \sin\theta} {\partial \over \partial
\theta}\bigg(\sin\theta {\partial \over \partial \theta}\bigg)
+{1\over \sin^2 \theta}{\partial^2  \over \partial \phi^2}\pm\mu^2
\bigg]\psi=0,
\end{eqnarray}
where before $\mu^2$ term the $'+'$ sign describes the phantom field
while the $'-'$ sign is for the quintessence field. Since the Vaidya
metric is spherically symmetric, above equation thus is separable
and the radial wave equation can be written as follows
\begin{eqnarray}\label{Radial3}
(1+\varepsilon_2){\partial ^2 \Psi \over \partial r_*^2} +2c
{\partial ^2 \Psi \over \partial v_* \partial r_*} +\varepsilon_1
{\partial \Psi \over \partial r_*}-V\Psi=0,
\end{eqnarray}
where
\begin{eqnarray}
\varepsilon_2&=&{f\over A}+2c {\partial r_* \over \partial v}-1, \\
\varepsilon_1&=&f A \bigg({1\over A}\bigg)'+f' +2cA {\partial \over
\partial v}{1\over A},\\ \label{V} V&=&{A f'\over r}+{\lambda A c^2\over
r^2}\mp c^2 \mu^2 A,
\end{eqnarray}
with $\frac{1}{A}=\frac{\partial r_*}{\partial
r}=1+\frac{1}{2\kappa(r-r_+)}$. Here in the effective potential $V$,
the $'-'$ sign is for the phantom field and the $'+'$ sign is for
the quintessence field.  For the convenience of later numerical
calculation, we adjust $\kappa$ in a way that Eq.(\ref{Radial3}) for
each multipole moment becomes the standard wave equation near the
horizon $r_+(v_0)$ \cite{rev}. The boundary condition at the horizon
reads
\begin{equation}
\mathop {\lim }\limits_{\begin{array}{l}
 r \to r_ \pm (v_0 ) \\
 v \to v_0 \\
 \end{array}} \varepsilon _2 = \mathop {\lim }\limits_{\begin{array}{l}
 r \to r_ \pm (v_0 ) \\
 v \to v_0 \\
 \end{array}} \varepsilon _1 = \mathop {\lim }\limits_{\begin{array}{l}
 r \to r_ \pm (v_0 ) \\
 v \to v_0 \\
 \end{array}} V = 0,
\end{equation}
which requires $\kappa={1\over 2r_+}$.

\begin{figure}[H]
\centering
\includegraphics[width=12cm]{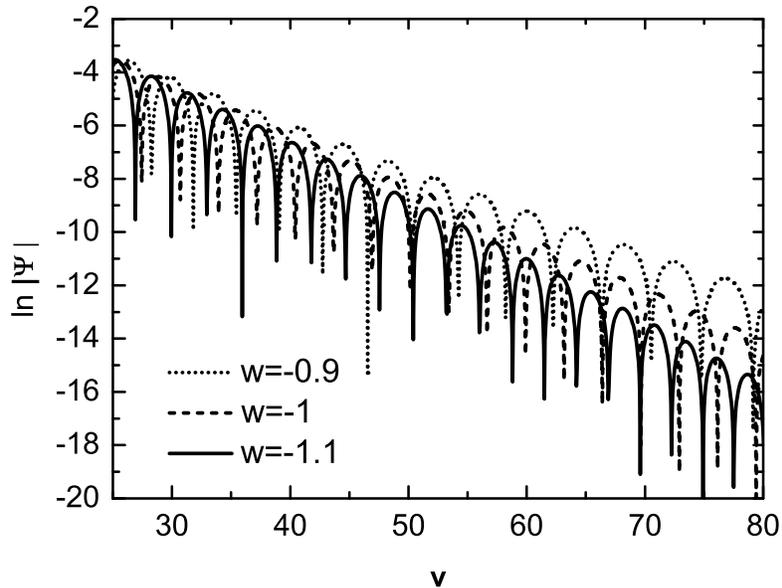}
\caption{\label{QNM} QNM behaviors of black holes absorbing dark
energy. In plotting the figure we have taken parameters $l=2$,
~$\rho_\infty=0.001$, ~$A\simeq 5.6$, $\mu=0.001$.}
\end{figure}

\begin{figure}[H]
\centering
\includegraphics[width=12cm]{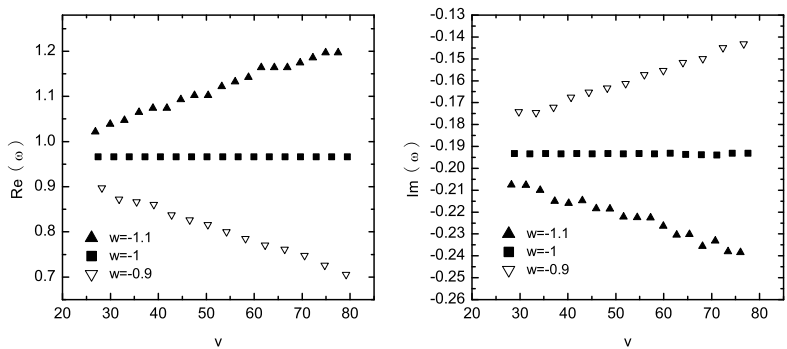}
\caption{\label{wrwi} The real and imaginary parts of quasinormal
frequencies when black hole absorbs different types of dark
energy. We have taken parameters $l=2$, ~$\rho_\infty=0.001$,
~$A\simeq 5.6$ and $\mu=0.001$ in plotting the figure }
\end{figure}

In order to simplify (\ref{Radial3}), we make a variable
transformation $u=u(r_*, v_*)$ and $v=v_*$, where the curve
$u(r_*,v_*)={\rm constant}$ is determined by the equation
\begin{equation}\label{dr*dv}
\frac{dr_*}{dv_*}= \frac{1+\varepsilon_2}{2c}\;.
\end{equation}
In addition, when $c=1$ and $\varepsilon_2\rightarrow 0$, we have
back the usual null coordinate $u\rightarrow v_*-2r_*$, similar to
the static case \cite{Shao:2004ws}. Using the new variables $u$ and
$v$, the wave equation changes to
\begin{eqnarray}\label{Wave}
\bigg[\big(1+\varepsilon_2\big)\bigg({\partial u\over \partial
r_*}\bigg)^2+2c{\partial u \over \partial r_*} {\partial u \over
\partial v}\bigg]{\partial ^2 \Psi \over \partial u^2}
+\bigg[\big(1+\varepsilon_2\big){\partial^2 u\over \partial r_*^2}+
2c {\partial^2 u \over \partial v \partial
r_*}+\varepsilon_1 {\partial u\over \partial r_*}\bigg]{\partial
\Psi \over \partial u}
+2c{\partial u \over
\partial r_*}{\partial^2 \Psi \over
\partial u \partial v}-V\Psi=0.
\end{eqnarray}
In the static limit,
$\varepsilon_2=\varepsilon_1=0$, and  $u=v_*-2r_*$, thus the radial
equation returns to its usual form \cite{rev}
\begin{eqnarray}
{\partial ^2 \Psi \over \partial u \partial v}+{V\over 4}\Psi=0.
\end{eqnarray}
Now for the Vaidya spacetime, $u(r_*,v_*)$ satisfies
\begin{eqnarray}
\begin{cases}
{(1+\varepsilon_2) {\partial u\over \partial r_*}+2c {\partial u
\over \partial v}=0}\\{(1+\varepsilon_2) {\partial^2 u\over \partial
r_*^2}+2c{\partial ^2 u\over \partial v\partial r_*}+\varepsilon_1
{\partial u\over \partial r_*}}=0,
\end{cases}
\end{eqnarray}
and Eq. (\ref{Wave}) simplifies to
\begin{eqnarray}\label{Wave1}
{\partial^2 \Psi \over \partial u \partial v}- \bigg(2c {\partial
u \over \partial r_*}\bigg)^{-1}V \Psi=0.
\end{eqnarray}

Numerical solution of (\ref{Wave1}) proceeds by first integrating
$u(r_*,v_*)$ according to (\ref{dr*dv}), and then using next the
finite difference method developed in \cite{Shao:2004ws} to solve
the wave equation in the Vaidya metric. Denoting $\Psi \to
\Psi_{i,j}=\Psi(u_i ,v_j) $, and
\begin{eqnarray}
{\partial \Psi \over \partial u}\to
{\Psi_{i+1,j}-\Psi_{i-1,j}\over \Delta u}, ~~~~~~{\partial \Psi
\over \partial v}\to {\Psi_{i,j+1}-\Psi_{i, j-1}\over \Delta v},
\end{eqnarray}
we can discrete (\ref{Wave1}) into
\begin{eqnarray}\label{discrete}
\Psi_{i-1,j+1} +\Psi_{i+1,j-1} -\Psi_{i-1,j-1}-\Psi_{i+1,j+1}
+\bigg(2c {\partial u \over \partial r_*}\bigg)^{-1}V \Psi_{i,j}=0.
\end{eqnarray}
Taking $\Psi_N=\Psi_{i+1,j-1}$, $\Psi_S=\Psi_{i-1,j-1}$,
$\Psi_E=\Psi_{i+1,j-1}$, $\Psi_W= \Psi_{i-1,j+1}$, and using
$\Psi_{i,j}= {\Psi_W +\Psi _E\over 2}$, we have
\begin{eqnarray}
\Psi_W+\Psi_E-\Psi_S-\Psi_N+\bigg(2c{\partial u\over \partial
r_*}\bigg)^{-1} {\Psi_E+\Psi_W \over 2}=0.
\end{eqnarray}

\begin{figure}[H]
\centering
\includegraphics[width=12cm]{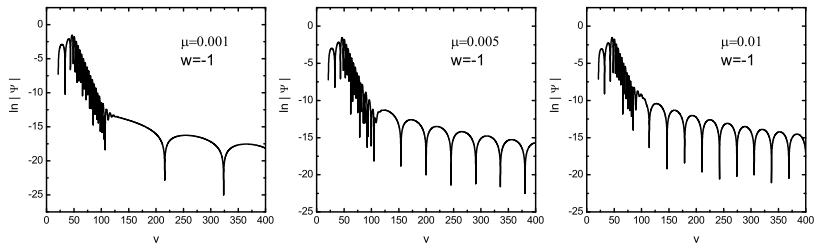}
\caption{\label{lateC} Late time evolution of the perturbation of
massive scalar field in the stationary spacetime. We have taken
$w=-1$, and different values of the field mass
~$\mu=0.001$,~$0.005$ and ~$0.01$ respectively.}
\end{figure}

\begin{figure}[H]
\centering
\includegraphics[width=12cm]{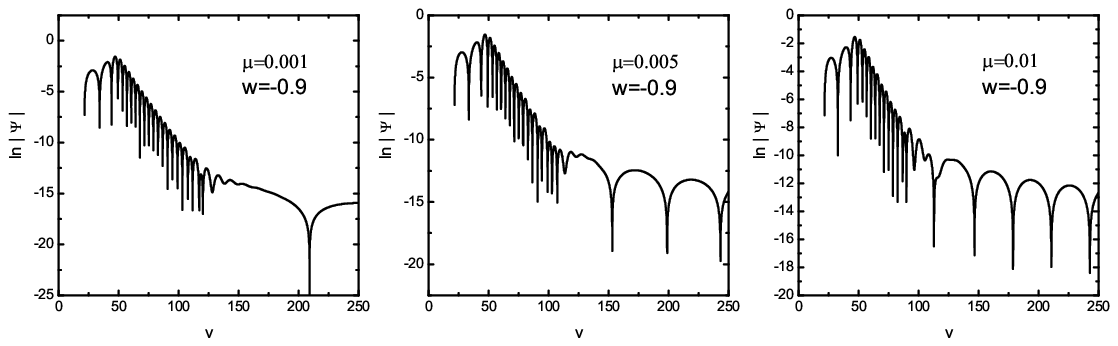}
\includegraphics[width=12cm]{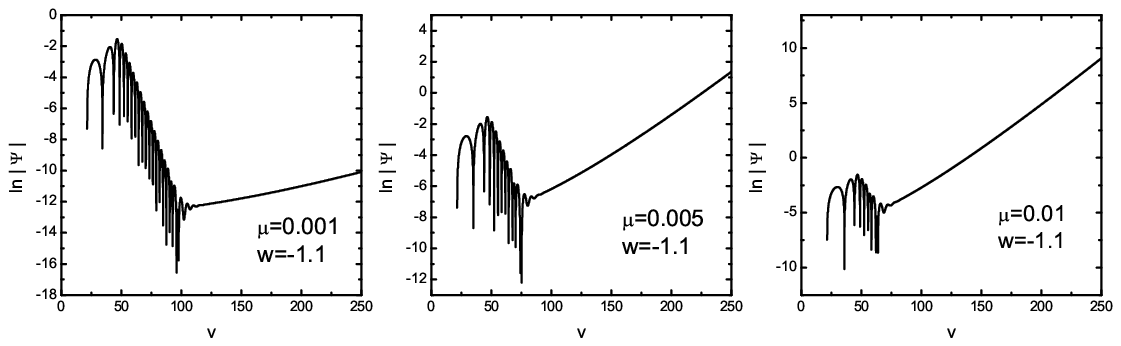}
\caption{\label{latePQ} Late time evolution of the perturbation in
around the black hole absorbing quintessence and phantom fields.
We have taken different values of the field mass as
~$\mu=0.001$,~$0.005$ and ~$0.01$ respectively.}
\end{figure}

Now we report our numerical results on the perturbations around the
black hole with accretion of different types of dark energy. Fig.1
illustrates the QNMs behavior. The solid line represents the case of
accretion of phantom type dark energy with $w=-1.1$, the dashed line
represents the dark energy being the cosmological constant, and the
dotted line shows the result of absorbing quintessence field with
$w=-0.9$. The details of quasinormal frequencies $\omega_R,
\omega_I$ as functions of $v$ are shown in Fig.2. When the black
hole absorbs the cosmological constant, its mass does not change so
that both the real and imaginary parts of quasinormal frequencies
remain constant, in agreement with the static black hole cases. The
black hole mass increases due to the accretion of quintessence
field. Differently from the stationary black hole case, we observed
that both the real part and the absolute value of the imaginary part
of quasinormal frequencies decrease with the time evolution. However
due to the accretion of phantom field, which causes the black hole
mass to decrease, both the real part and the absolute value of the
imaginary part of quasinormal frequencies increase with time.
Comparing the values of the imaginary parts of QNM, we found that
$|\omega_I|$ is smaller for black holes absorbing quintessence than
phantom field. This explains what we observed in Fig.1, in which the
effect of perturbation can last longer in the black hole background
due to the accretion of quintessence type dark energy.

We have also investigated the late time tail behavior of the
perturbation in the black hole due to the accretion of dark
energy. If the dark energy is cosmological constant, the result is
shown in Fig.3. Since the black hole mass do not change in this
case, our result gives the objective picture of the late time tail
behavior for the massive scalar field perturbation in the
stationary black hole background. Differently from massless scalar
field perturbation, we observed oscillations in the tail of
massive scalar field perturbation. The frequency of the
oscillation increases with the value of mass of scalar field
$\mu$, which confirms the argument proposed in \cite{mass}.

\begin{figure}
\centering
\includegraphics[width=12cm]{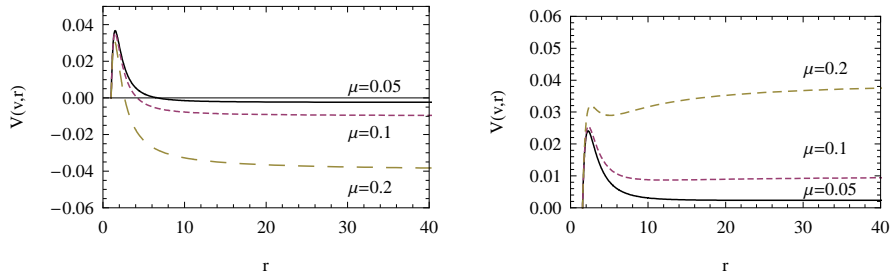}
\caption{\label{V1} Behaviors of the effective potential $V(r)$ at
the moment $v=100$ for different values of $\mu$ for the accretion
of phantom and quintessence respectively. In plotting the figure, we
have taken $l=0$, ~$\rho_\infty=0.001$, ~$A\simeq 5.6$.}
\end{figure}

The late time tails of perturbations around black holes absorbing
quintessence field and phantom field are shown in Fig.4. The
accretion of quintessence field has the similar oscillatory decay
behavior in the late time tail as that observed for black hole
absorbing dark energy with $w=-1$. Interestingly, the accretion of
phantom onto the black hole presents us completely different
behavior in the late time tail. Instead of decay, we saw the growing
modes in the tail. The same phenomenon has also been observed for
static black holes surrounded with phantom type DE
\cite{Chen:2008ru}. Here we found that the growing appears earlier
and faster when $\mu$ is bigger. The growing modes will make the
black hole unstable. Thus from the tail of the perturbation we saw
that it is not peaceful for the black hole to disappear by the Big
Rip due to the accretion of phantom energy as argued in \cite{bh}.
In the later moment when the black hole has accumulated enough
phantom energy, it will ``explode". To find physical reason for this
``explosion", let us examine the behavior of the effective potential
in the wave equation. Since the black hole mass will change due to
the accretion of phantom or quintessence fields, the effective
potential is time-dependent. Choosing $v=100$, the behaviors of the
potential are shown in Fig.5. For the accretion of phantom, the
effective potential outside the black hole approaches a negative
value, which is different from the case with accretion of the
quintessence. Actually this can be seen from the expression of the
potential (\ref{V}), whose  behavior does not change with the time
evolution. The existence of the negative part in the effective
potential can form true bound state which leads to the growing modes
\cite{Wang:2001tk}.

In summary we have investigated perturbations around black holes
absorbing dark energy. It is well known that the mass of a black
hole changes due to the accretion of dark energy of quintessence and
phantom types\cite{bh}. The study of perturbation in this black hole
background is more complicated than in the usual stationary
situation. Using the formalism developed in \cite{Shao:2004ws}, we
have observed different QNM behaviors for the accretion of different
types of dark energy onto the black hole. The QNM results discussed
here are sufficient to illustrate the possibility to distinguish
whether dark energy lies above or below the cosmological constant
boundary $w=-1$ in the future observation by exploiting the
perturbations around black holes. However since the change rate of
the mass due to the accretion of dark energy is extremely slow, to
extract very accurate information of $w$ from gravitational waves
emitted from black holes while avoiding background noise still
remains a challenge.

The late time tail behavior of the perturbation around black hole
absorbing dark energy presents us interesting results. Instead of
the anticipated oscillatory decay in the background of black hole
absorbing cosmological constant and quintessence, the accretion of
phantom energy onto the black hole exhibits a growing mode in the
perturbation. This growing tail tells us that due to the accretion
of phantom energy the black hole does not disappear peacefully as
argued in \cite{bh}. Instead, the black hole will explode after getting
enough phantom energy. The result is consistent with the
Big Rip scenario, i.e, all of the bound objects are to be torn
apart in the presence of the phantom dark energy.

\acknowledgments{This work was supported in part by NNSF of China,
Shanghai Education Commission and Shanghai Science and Technology
Commission. The work of C.-Y. L. was supported in part by the
National Science Council under Grant No. NSC-95-2112-M-259-003}

\end{document}